\documentclass{vgtc}                          % final (conference style)
\graphicspath{{figures/}{pictures/}{images/}{./}} % where to search for the images

\usepackage{times}                     % we use Times as the main font
\usepackage{algorithm}                 % 슈도코드
\usepackage[noend]{algpseudocode}
\usepackage{url}
\usepackage{hyperref}
\usepackage{graphicx}
\usepackage{xcolor}
\usepackage{kotex}
\usepackage{subfig}
\usepackage{tikz}
\usepackage{booktabs}

\newcommand{\ourtool}{\texttt{MetaScanner}}

\onlineid{1016}

\vgtccategory{Systems}

\title{Illusion Worlds: Deceptive UI Attacks in Social VR\thanks{This work was presented as a poster at IEEE VR 2025. It will be included in the official conference proceedings.}}

\author{Junhee Lee\thanks{e-mail: hunroung@kw.ac.kr}\\ %
    \scriptsize Kwangwoon University %
\and Hwanjo Heo\thanks{e-mail: hwanjo@etri.re.kr}\\ %
     \scriptsize ETRI %
\and Seungwon Woo\thanks{e-mail: seungww@etri.re.kr}\\ %
     \scriptsize ETRI
\\
\and Minseok Kim\thanks{e-mail: rhdtls1000@kw.ac.kr}\\
    \scriptsize Kwangwoon University
\and Jongseop Kim\thanks{e-mail: joseab5965@kw.ac.kr}\\
    \scriptsize Kwangwoon University
\and Jinwoo Kim\thanks{e-mail: jinwookim@kw.ac.kr (corresponding author)}\\ 
    \scriptsize Kwangwoon University
}
\abstract{
Social Virtual Reality (VR) platforms have surged in popularity, yet their security risks remain underexplored. This paper presents four novel UI attacks that covertly manipulate users into performing harmful actions through deceptive virtual content. Implemented on VRChat and validated in an IRB-approved study with 30 participants, these attacks demonstrate how deceptive elements can mislead users into malicious actions without their awareness. To address these vulnerabilities, we propose \ourtool{}, a proactive countermeasure that rapidly analyzes objects and scripts in virtual worlds, detecting suspicious elements within seconds.
}


\begin{document}



\maketitle

\section{Introduction}

As Virtual Reality (VR) continues to grow in popularity, it has become a prime target for adversaries exploiting its immersive interactions, leading to significant security risks such as privacy invasion and user manipulation. Prior research has highlighted threats, including misleading users through distorted virtual environments~\cite{lee2021adcube,tseng2022dark} and exposing personal data via head-tracking and performance metrics~\cite{zhang2023head}. Despite these findings, the security risks of social VR platforms like VRChat remain largely unexamined. With their increasing adoption, addressing these vulnerabilities is both timely and essential.

This paper explores the fundamental vulnerability of social VR platforms: the ability for users to create \emph{deceptive virtual content}. Thus, adversarial content creators can exploit (i)~virtual worlds, (ii)~avatars, and (iii)~user interactions, embedding harmful logic that appears legitimate through social engineering techniques. As users are encouraged to explore diverse virtual content to fulfill social needs~\cite{sykownik2021most}, these attacks effectively attract participation while remaining difficult to detect. Specifically, we propose four novel UI attacks: (i)~\emph{Object Clickjacking}, capturing inputs and redirecting them to unintended content; (ii)~\emph{Denial-of-Raycasting}, blocking user interactions via invisible objects; (iii)~\emph{Object-in-the-Middle}, exfiltrating sensitive information like passwords; and (iv)~\emph{Avatar Quishing}, deceiving users with malicious QR codes embedded in avatars. Our IRB-approved user study with 30 participants demonstrates the effectiveness of these attacks within a proof-of-concept virtual world on VRChat.

To mitigate these risks, we introduce \ourtool{}, a proactive countermeasure designed to rapidly detect suspicious objects and scripts, including invisible elements and malicious URLs, within virtual content. Our evaluation of \ourtool{} across 38 collected virtual environments demonstrates its effectiveness in identifying deceptive virtual content with minimal processing time.

% \noindent\textbf{Contributions.} Our contributions are summarized as follows:
% \begin{itemize}
%     \item We introduce four novel UI attacks that can be covertly mounted by an adversarial content creator. To the best of our knowledge, this is the first work to exploit the social VR ecosystem, in which users can become content creators with potentially malicious intent.
%     \item We demonstrate the feasibility of our proposed attacks by implementing and showcasing our custom virtual world on VRChat, the most popular social VR platform, and present an extensive user study that highlights their effectiveness.
%     \item We propose an effective and lightweight defense system \ourtool{} that proactively detects prospective UI attacks before an adversary-created virtual worlds go online.
% \end{itemize}
\section{Attacks}
\label{section:ui_attacks}

We propose four novel attacks targeting UIs in social VR platforms. These attacks enable adversaries to execute malicious actions stealthily and persistently, leveraging the immersive nature of VR to exploit user interactions.

\subsection{Object Clickjacking}
\label{subsec:clickjacking_design}

\begin{figure}[h]
    \centering
    \includegraphics[width=.9\linewidth, trim={0 1cm 0 0}]{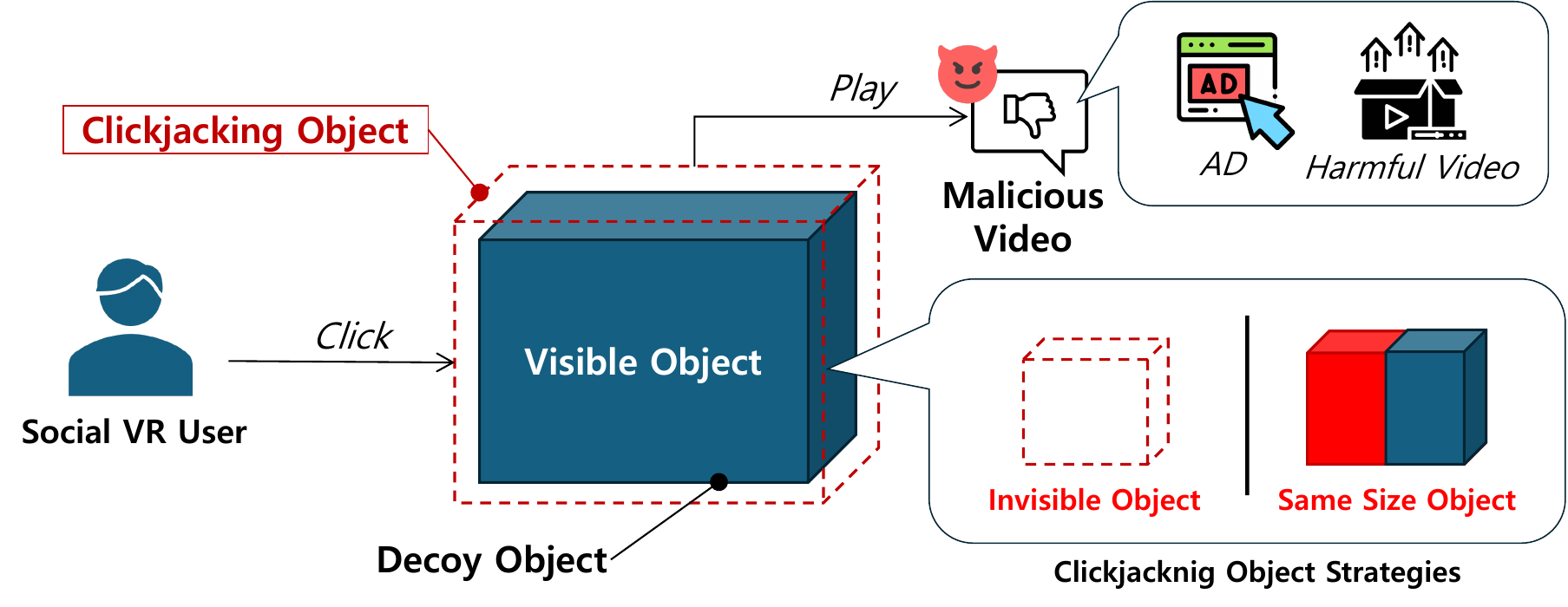}
    \caption{An object clickjacking attack in which the \emph{clickjacking object} intercepts user input intended for the \emph{decoy object}, triggering a malicious action.}
    \label{fig:attack1}
    \vspace{-0.1in}
\end{figure}

In social VR, adversaries take advantage of immersive interactions using strategies such as deploying benign and malicious objects in the same position, exploiting rendering order inconsistencies (\emph{same size object}), or overlaying transparent objects similar to traditional clickjacking (\emph{invisible object}). As shown in Figure~\ref{fig:attack1}, these techniques can initiate harmful events without the user's consent, such as redirecting them to illegal ads or malicious videos.

\subsection{Denial-of-Raycasting}
\label{subsec:raycasting}

\begin{figure}[h]
    \centering
    \includegraphics[width=.9\linewidth, trim={0 1cm 0 0}]{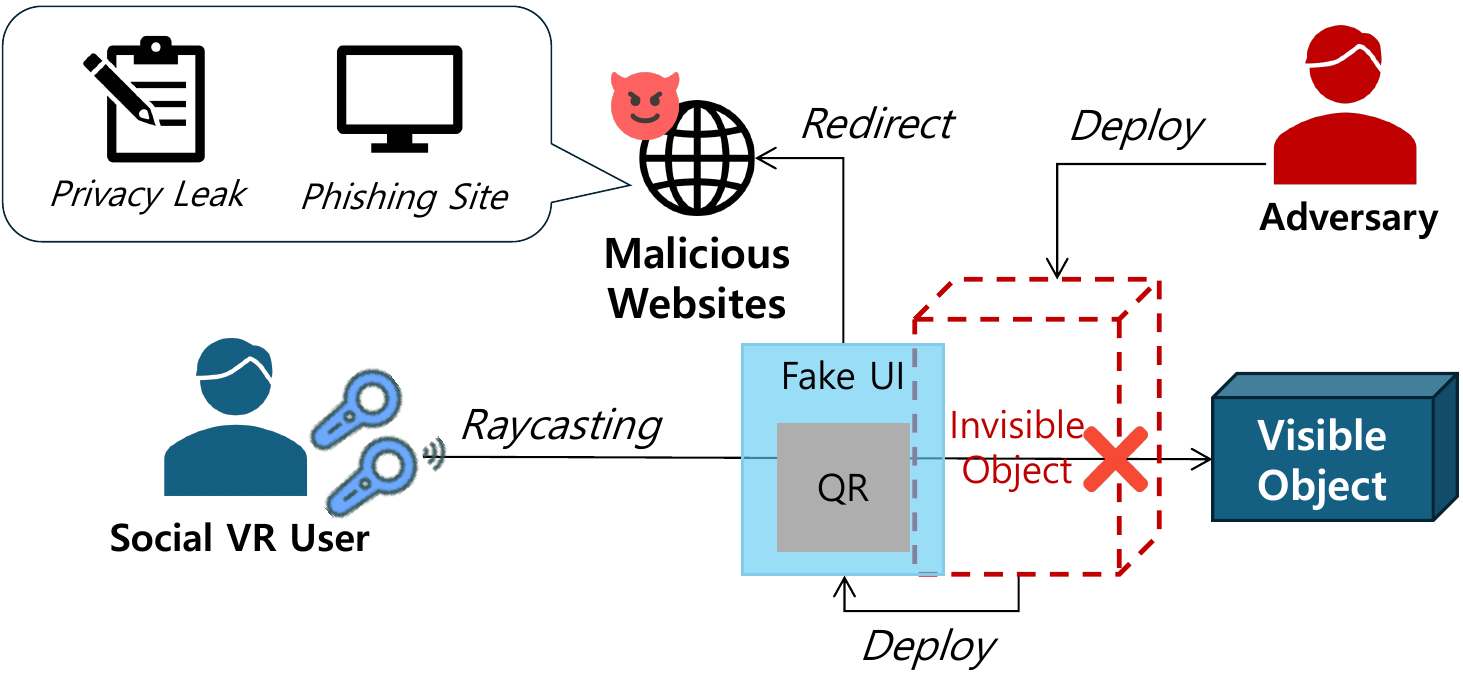}
    \caption{A denial-of-raycasting attack, where an \emph{invisible object} blocks the user's raycasting path.}
    \label{fig:attack2}
    \vspace{-0.1in}
\end{figure}

In denial-of-raycasting attacks (Figure~\ref{fig:attack2}), adversaries block user interactions with visible objects by placing invisible objects along raycasting paths. This can mislead users into believing their controller is malfunctioning. Adversaries may also present fake troubleshooting guides containing phishing QR codes that redirect users to malicious sites.

\vspace{-0.05in}
\subsection{Object-in-the-Middle}
\label{subsec:oitm}
\vspace{-0.05in}

\begin{figure}[h]
    \centering
    \includegraphics[width=.9\linewidth, trim={0 1cm 0 0}]{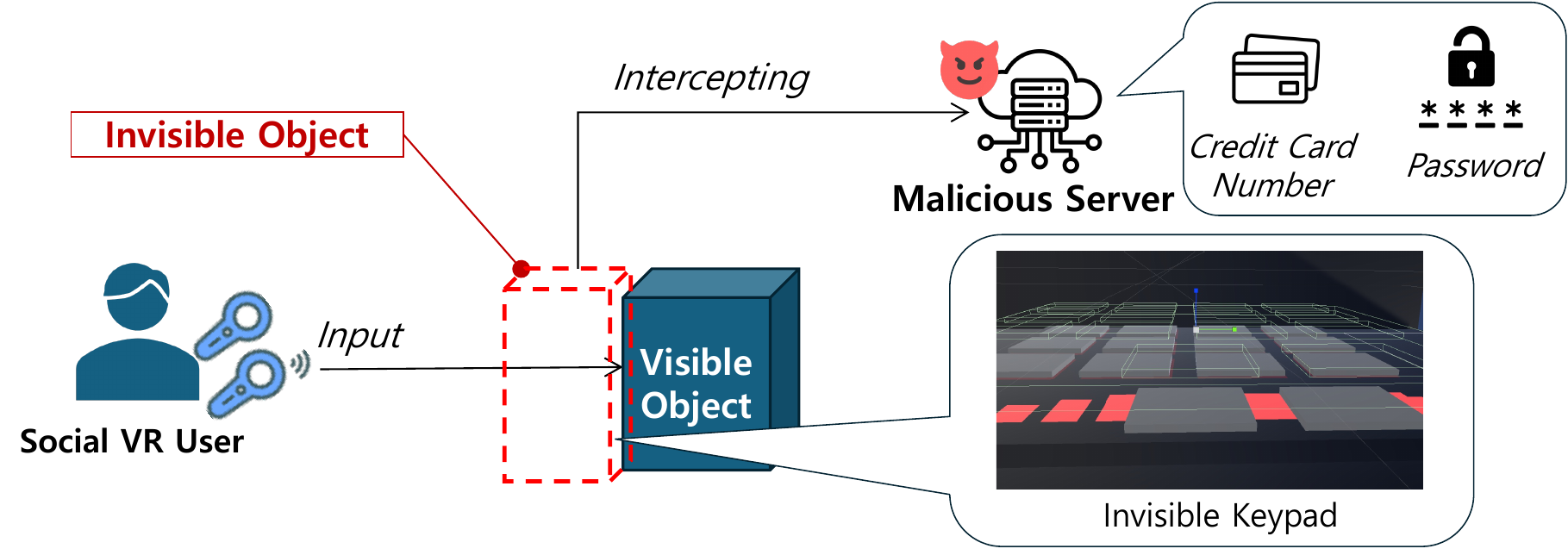}
    \caption{An object-in-the-middle attack, where an \emph{invisible object} intercepts keystrokes on a virtual keypad.}
    \label{fig:attack3}
    \vspace{-0.1in}
\end{figure}

Adversaries in object-in-the-middle attacks (Figure~\ref{fig:attack3}) intercept sensitive inputs, such as credit card numbers, using invisible objects over virtual keypads. Captured data can either be exfiltrated to external servers or stored within hidden areas of the virtual world for later retrieval, making the attack both stealthy and persistent.

\vspace{-0.05in}
\subsection{Avatar Quishing}

\begin{figure}[h]
    \centering
    \includegraphics[width=.9\linewidth, trim={0 1cm 0 0}]{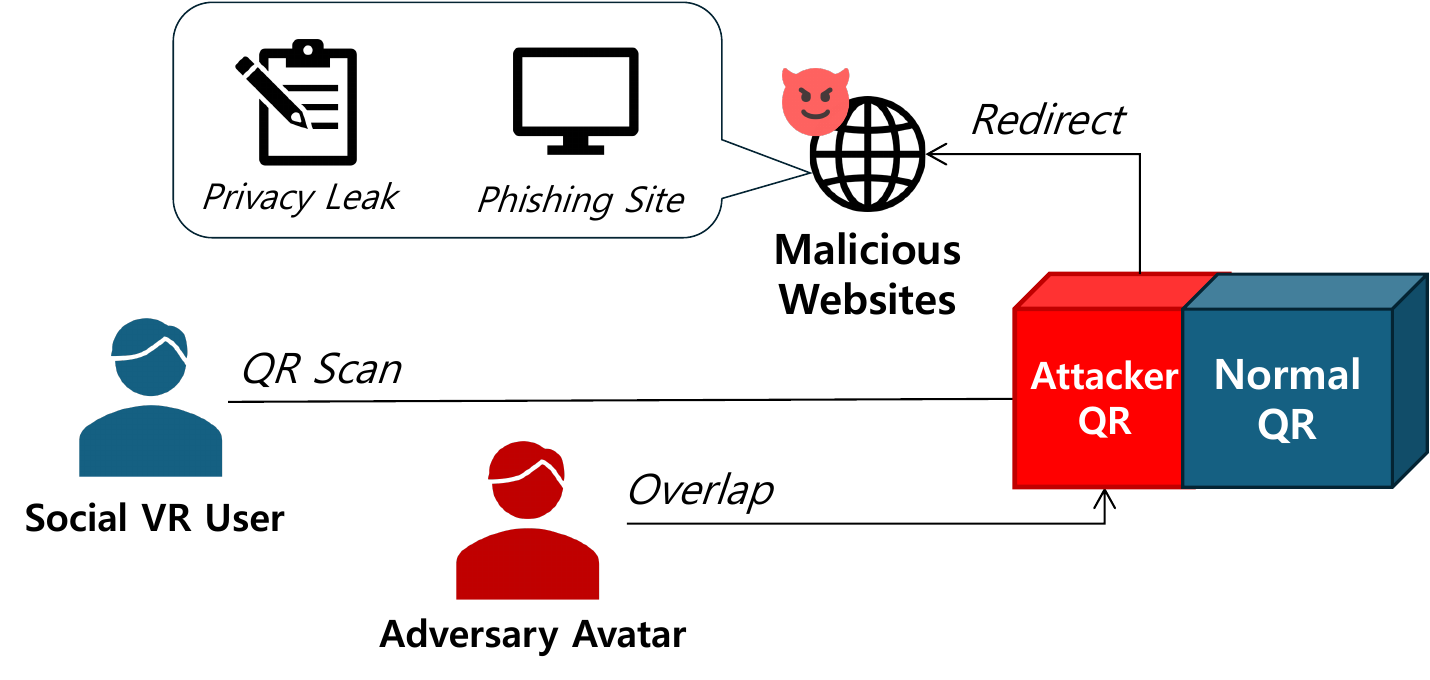}
    \caption{An avatar quishing attack where a \emph{malicious QR code} redirects user input to a malicious website.}
    \label{fig:quishing_attack1}
    \vspace{-0.1in}
\end{figure}

In avatar quishing (Figure~\ref{fig:quishing_attack1}), adversaries embed malicious QR codes into avatars. These codes, often detached from the avatar’s body, can covertly overlap legitimate QR codes, redirecting users to harmful websites. Unlike embedding codes in static virtual objects, leveraging avatars allows adversaries to infiltrate multiple virtual worlds, significantly complicating detection efforts.
\section{Results}
\label{section:user_study}

% \begin{figure}[t]
%     \centering
%     \subfloat{
%         \centering
%         \includegraphics[width=.46\linewidth]{figures/1-1.pdf}
%         \label{fig:1}
%     }
%     \hfill
%     \subfloat{
%         \centering
%         \includegraphics[width=0.46\linewidth]{figures/2-1.pdf}
%         \label{fig:2}
%     }
%     \vspace{-0.2in}
%     \subfloat{
%         \centering
%         \includegraphics[width=0.46\linewidth]{figures/3-1.pdf}
%         \label{fig:3}
%     }
%     \hfill
%     \subfloat{
%         \centering
%         \includegraphics[width=0.46\linewidth]{figures/4-1.pdf}
%         \label{fig:4}
%     }
%     \caption{Key question results from each attack experiment}
%     \label{fig:1}
%     \vspace{-0.2in}
% \end{figure}

Our IRB-approved user study for 30 participants demonstrates the effectiveness of all four proposed attacks in exploiting user vulnerabilities within social VR platforms. In \emph{Object Clickjacking}, 83.3\% of participants did not notice differences between the buttons due to their identical appearance and similar feedback. In \emph{Denial-of-Raycasting}, 86.7\% of participants failed to detect any changes after triggering the attack, and alarmingly, some participants expressed a willingness to scan a fake QR code displayed during the attack.
In \emph{Object-in-the-Middle}, all participants (100\%) failed to notice that their inputs were being intercepted by a transparent object. Similarly, in \emph{Avatar Quishing}, all participants (100\%) were unable to distinguish between malicious and benign QR codes, with several admitting they had never questioned the authenticity of a QR code before scanning. These findings underscore the critical vulnerabilities of users to deceptive UI-based attacks in social VR platforms.

\section{Defense}
\label{section:metascanner}

\begin{figure}[t]
    \centering
    \includegraphics[width=.95\linewidth]{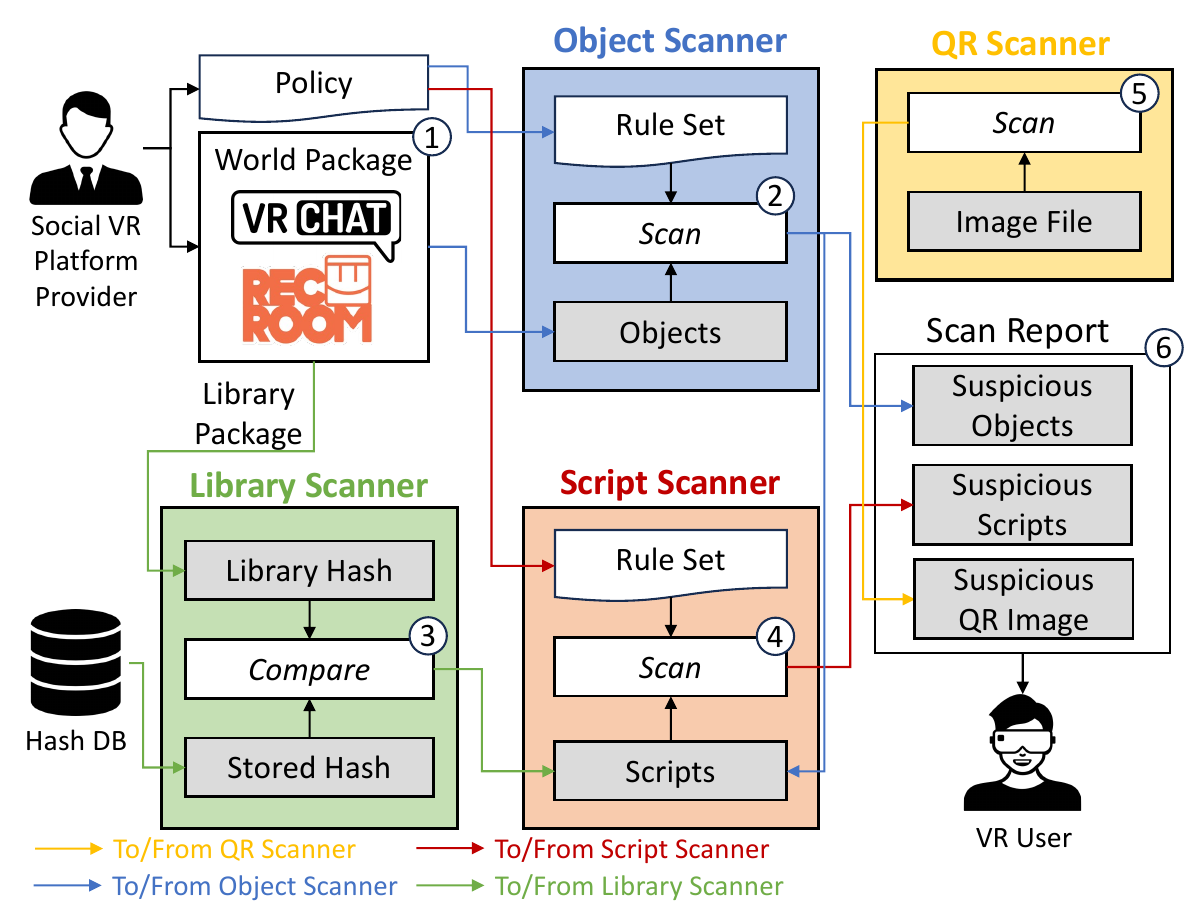}
    \caption{An architectural overview of \ourtool{}.}
    \label{fig:metascanner}
    \vspace{-0.20in}
\end{figure}

We posit that the proposed attacks originate from deceptive content that adversarial creators can deploy on social VR platforms. To address this issue, we present \ourtool{}, a static analysis tool designed to identify suspicious objects and scripts in Unity-based social VR platforms (Figure~\ref{fig:metascanner}). \ourtool{} consists of four submodules: (i) an Object Scanner, (ii) a Library Scanner, (iii) a Script Scanner, and (iv) a QR Scanner. It translates platform provider-defined \emph{policies} into actionable \emph{rule sets}, enabling the sequential analysis of objects, libraries, scripts, and QR codes to detect malicious elements. Finally, \ourtool{} produces a detailed report of detected threats and issues user warnings (Figure~\ref{fig:metascanner}). Through implementation and evaluation in 38 collected virtual worlds, we demonstrate the tool's effectiveness in detecting suspicious content with low processing overhead.

% \noindent\textbf{Object Scanning.} \ourtool{} detects \emph{suspicious objects}, such as transparent or overlapping objects, often used for clickjacking or object-in-the-middle attacks. It identifies object properties (e.g., transparency settings) and associated scripts, passing this data to the script scanner for further analysis.

% \noindent\textbf{Library Scanning.} Libraries from Unity and third-party developers are prone to supply chain attacks~\cite{ladisa2023sok}. \ourtool{} ensures library integrity by comparing their hash values against a database of trusted libraries. Any discrepancies trigger further script analysis.

% \noindent\textbf{Script Scanning.} The script scanner uses static analysis to detect harmful APIs, such as \texttt{Application.OpenURL} and \texttt{System.Net.Sockets}, which can be exploited for information leakage or malicious actions.

% \noindent\textbf{QR Scanning.} The QR scanner identifies suspicious QR codes in image files within the virtual world's assets. Detected URLs are flagged, and users are notified of potential risks.

% This streamlined design ensures efficient detection of threats across large-scale virtual environments.

\vspace{-0.05in}
\section{Conclusion and Future Work}
This paper presents novel UI attacks on social VR platforms, highlighting how adversarial content creators embed stealthy attacks within virtual worlds and avatars. Future work will focus on validating the feasibility of these attacks in real-world social VR environments, such as VRChat, through more realistic experiments. Additionally, we aim to enhance the evaluation of \ourtool{} in these environments to further demonstrate its effectiveness.

\acknowledgments{
This work was partially supported by Institute of Information \& communications Technology Planning \& Evaluation (IITP) [RS-2023-00215700, Trustworthy Metaverse: blockchain-enabled convergence research] and the National Research Foundation of Korea (NRF) (No.~RS-2024-00457937) grant funded by the Korea government (MSIT).
}
\vspace{-0.1in}

\bibliographystyle{abbrv-doi}
\bibliography{main}
\end{document}